\documentclass[aps,prb,twocolumn,amsmath,amssymb,superscriptaddress,floatfix,scrartcl,eqsecnum]{revtex4-1}

\usepackage{amssymb}
\usepackage{color}
\usepackage{graphicx}
\usepackage{mathrsfs}
\usepackage[unicode=true,pdfusetitle,
bookmarks=false,colorlinks=true,citecolor=blue,urlcolor=blue,linkcolor=red]{hyperref}

\usepackage{bm}
\usepackage{braket}
\usepackage{soul}
\usepackage{float}
\usepackage{amsmath,amssymb,mathrsfs,bm,setspace}
\usepackage{graphicx}
\usepackage[tight]{subfigure}
\usepackage{color}

\usepackage{graphicx}
\usepackage{amsmath}
\usepackage{amssymb}
\usepackage{bm, bbm}      
\usepackage{color}
\usepackage{srcltx}

\usepackage{slashed}

\def\pd{\partial}
\def\=d{\, {\buildrel \rm def  \over =} \,}

\def\sqr#1#2{{\vcenter{\vbox{\hrule height.#2pt \hbox{\vrule width.#2pt height#1pt \kern#1pt \vrule width.#2pt}\hrule height.#2pt}}}}

\def\beq#1{\begin{equation} \label{#1}}
\def\eeq{\end{equation}}
\def\ben{\begin{equation*}}
\def\een{\end{equation*}}
\def\bequa{\begin{eqnarray}}
\def\eequa{\end{eqnarray}}

\def\bf#1{\bm{#1}}

\def\Tr{\mathop{\mathrm{Tr}}}

\def\tr{\text{tr}}
\def\phie{\phi_E}

\newcommand\eea{\end{eqnarray}}
\newcommand\bea{\begin{eqnarray}}

\def\beq{\begin{equation}}
\def\eeq{\end{equation}}

\def\al{\alpha}

\def\ren{R\'{e}nyi }
\def\mue{{\mu_E}}

\newcommand{\be}{\begin{equation}}
\newcommand{\ee}{\end{equation}}
\newcommand{\ba}{\begin{align}}
\newcommand{\ea}{\end{align}}
\newcommand{\bg}{\begin{gather}}
\newcommand{\eg}{\end{gather}}
\newcommand{\bseq}{\begin{subequations}}
\newcommand{\eseq}{\end{subequations}}

\begin{document}

\title{Charged Topological Entanglement Entropy}

\author{Shunji Matsuura}
\affiliation{Niels Bohr International Academy and Center for Quantum Devices,
Niels Bohr Institute,
Copenhagen University,
Blegdamsvej 17,
Copenhagen, Denmark}
\affiliation{Yukawa Institute for Theoretical Physics, Kyoto University, Kyoto, Japan }
\author{Xueda Wen}
\affiliation{Institute for Condensed Matter Theory and Department of Physics, University of Illinois at Urbana-Champaign, 1110 West Green St, Urbana IL 61801 }
\author{Ling-Yan Hung}
\affiliation{Department of Physics and Center for Field Theory and Particle Physics, Fudan University, 220 Handan Road, 200433 Shanghai, China}
\affiliation{Collaborative Innovation Center of Advanced Microstructures, Fudan University, 220 Handan Road, 200433 Shanghai, China}
\author{Shinsei Ryu}
\affiliation{Institute for Condensed Matter Theory and Department of Physics, University of Illinois at Urbana-Champaign, 1110 West Green St, Urbana IL 61801 }

\begin{abstract}

A charged entanglement entropy is a new measure which probes quantum entanglement between different charge sectors.
We study symmetry protected topological (SPT) phases in 2+1 dimensional space-time by using this charged entanglement entropy.
SPT phases are short range entangled states without topological order and 
hence cannot be detected by the topological entanglement entropy.
We demonstrate that the universal part of the charged entanglement entropy is non-zero for non-trivial SPT phases
and therefore it is a useful measure to detect short range entangled topological phases.
We also discuss that the classification of SPT phases based on the charged topological entanglement entropy 
is related to that of the braiding statistics of quasiparticles.

\end{abstract}
\maketitle

{\hypersetup{linkcolor=black} \tableofcontents}

\newpage

\section{Introduction}

One feature that distinguishes quantum physics from classical physics is its non-locality:
two spatially separated regions can have non-trivial correlations.
It has been recognized that this non-locality may play a central role in characterizing different quantum phases of matter.
While Landau's symmetry breaking paradigm focuses on the behavior of local order parameters,
topological phases focus on the global structure of quantum entanglement.
Therefore, topological phases are genuinely of quantum origin.
Since the discovery of the quantum Hall effect \cite{FQH}, 
it is recognized that there can be distinct topological phases, 
which have different patterns of quantum entanglement \cite{CGW,WenInt,WenNiu}.

An essential idea to topologically classify different gapped phases of quantum matter 
is to ask whether different ground states
are connected to each other by a continuous deformation.
Namely, 
when two states are transformed into each other by 
local unitary transformations with a finite depth,
they belong to the same phase and are topologically equivalent.
On the other hand, 
if they are not, they belong to different phases.  
For a latest comprehensive discourse, see \cite{wenbook}.

If a state cannot be connected to a topologically trivial states, i.e., a product state, 
by any local unitary transformation, 
it is called a long range entangled state.
For example, with this definition, 
integer quantum Hall states (Chern insulators), and fractional quantum Hall states 
are long range entangled states.
Most of the states with long range entanglement have a specific scaling form of the entanglement entropy.
As suggested from the area law of the entanglement entropy, 
the dominant contribution to the entropy is from the vicinity of the entangling surface.
In addition to this short range non-universal contribution, 
long range entangled states have topological contribution,
which does not depend on the shape of the subsystem.
For a long range entangled state in (2+1) dimensions, which is placed on a plane,
when we choose a disk region of size $L$ as a subsystem, 
where $L$ is much larger than other scales such as the correlation length,
the entanglement entropy behaves as
\cite{KitaevPreskill, wenx}
\bea
{S_{EE}}=aL-\gamma+\cdots.
\eea
The first term, the area law term, arises from short range physics near the entangling surface
and the coefficient $a$ is non-universal in the sense that it depends on the cut-off.
The subleading term $\gamma$ is the topological entanglement entropy and
measures the total quantum dimension of a topological phase.

On the other hand,
gapped states which can be continuously connected to a product state 
are called short range entangled states\cite{SRE}.
In this sense, all gapped short-range entangled states belong to the same phase and completely trivial.
However, if we restrict the type of local unitary transformations, 
by imposing a symmetry, say, 
there can be distinct phases even within the short range entangled states.
For instance, in 3+1 dimensional non-interacting fermion systems, 
there is no states with long range entanglement 
(i.e., there is no analogue of the integer quantum hall effect in 3+1 dimensions).
However, if we impose time-reversal symmetry,
there are states which are topologically distinct from trivial insulators.
This is an example of topological insulators and superconductor 
\cite{HasanKane,QiZhang},
which are classified by K-theory\cite{Kitaev-K,Schnyder-K}.
More generally, if a state is connected to a product state by general local unitary transformations
but not by
local unitary transformations with certain symmetry, it is called a symmetry protected topological phase 
(SPT phases).
Since the topological entanglement entropy is zero for all short-range entangled states, 
it is not a useful measure for distinguishing SPT phases.

In this paper, 
we consider the "grand canonical entanglement entropy"\cite{Belin:2013uta, rotating ent,charge-refs},
which have been discussed, e.g., in the context of the gauge/gravity duality.
In short, the grand canonical entanglement entropy is an extension (a generalization) of 
the ordinary (von-Neumann and \ren) entanglement entropy by introducing
the conjugate variables that couple to conserved quantities,
such as charges and (angular) momentum, etc., 
in the subregion of our interest.
(See below for more detailed definitions.)
In particular, 
the grand canonical entanglement entropy 
defined in the presence of a potential 
(the "entanglement chemical potential") that is conjugate to a conserved charge
is called the charged entanglement entropy. 
Similarly,
the grand canonical entanglement entropy 
defined in the presence of the
conjugate variable that couples to momentum
is called the shifted entanglement entropy. 

In the context of the gauge/gravity duality, 
the grand canonical entanglement entropy can be naturally motivated/introduced as follows.
In the gauge/gravity duality, the entanglement entropy is measured by the thermal entropy of black holes 
whose horizon ends on the entangling surface at the boundary of AdS space-time.
This idea was used in \cite{Headrick:2010zt,Casini:2011kv} to prove the Ryu-Takayanagi formula\cite{rt0}
of the holographic entanglement entropy.
This class of black holes, dual to the entanglement entropy, has non-compact horizons. 
Nevertheless they satisfy the standard thermodynamical laws. 
In particular, one can consider charged and rotating black holes\cite{caution on RBH}, 
which will follow the grand canonical ensemble.
It is then natural to ask what 
the meaning of those charged and rotating black holes is, in the context of the entanglement entropy.
It is shown in \cite{Belin:2013uta} that those grand canonical entanglement entropies measure 
the charge/angular momentum fluctuations across the entangling surface 
that partitions the system into two regions.
It has been shown that there is a phase transition in \ren entropy
in the presence of a scalar and an electric charge\cite{Belin:2014mva}.

The purpose of this paper is to demonstrate 
another utility of the grand canonical entanglement entropy in condensed matter systems,
-- SPT phases.
In fact, since the presence of symmetries is a prerequisite 
for both SPT phases and the charge entanglement entropy
it is natural to expect that the latter is useful to study the former. 
We show that they can distinguish different short range entangled states;
In particular, we show that for (2+1)-dimensional SPT phases, 
the charged entanglement entropy $S_{EE}(\mue,\phie)$ behaves as
\bea
S_{EE}(\mue,\phie)=aL +\gamma_c+\cdots,
\eea
where the universal part is
\bea
\gamma_c=  \mathscr{C} (2\pi i \mue\phie)
\label{ctee}
\eea
Here, $\mue$ and $\phie$ are grand canonical potentials that couples to 
symmetry ``charges" of SPT phases. 
$\mathscr{C}$ is a topological number characterizing the SPT phase.
Notice that $\gamma_c$ is pure imaginary; it is a phase factor of the partition function on a replica space.
As such, one may think it is defined modulo $2\pi$. However, this identification has a meaning only if $\mue$ and $\phi$
are physically determined. In our case, they are just parameters and one could take $\mue$ as small as one wants so that
$\gamma_c$ is always between $0$ and $2\pi$ for any value of $\mathscr{C}$. 
In fact, a physical information is in $\mathscr{C}$ rather than $\gamma_c$.
As we will see, the identification of $\mathscr{C}$ is not by the periodicity of $\gamma_c$ but by a different physical reasoning.

The organization of this paper is as follows.
Sec.\ \ref{def sec of charged chem}, we review the charged and shifted entanglement entropies.
Sec.\ \ref{bulk story}, we calculate the charged entanglement entropies from bulk point of view.
Sec.\ \ref{sec CFT CI}, we rederive the charged entropies by using the edge theories.
Sec.\ \ref{SPT story}, we apply the charged entanglement entropies to SPT phases.
Sec.\ \ref{Shifted bulk story}, we compute the shifted entanglement entropy in the bulk theory.



\section{Grand canonical entanglement entropies and flux operators}
\label{def sec of charged chem}

\subsection{Charged entanglement entropies}

In Refs.\ \onlinecite{Belin:2013uta, rotating ent,charge-refs,mom-pol}, 
the entanglement entropy is generalized by introducing chemical potentials and angular potentials.
In this section, we review the definitions and basic properties.

Let us consider a quantum system and divide it spatially into two parts, 
a subsystem $V$ and its complement $\bar{V}$.
The reduced density matrix of the subsystem $V$ is given by tracing out the Hilbert space 
of the subsystem $\bar{V}$:
\bea
\rho^{\ }_{V}={\tr}_{\bar{V}}\rho,
\label{rhoV}
\eea
where $\rho$ is the density matrix corresponding to the pure ground state of the total system.
From the positivity and the hermiticity of the reduced density matrix, 
one can define an entanglement Hamiltonian $H_E$ as
\bea
\rho^{\ }_V={e^{- \beta H_E}\over Z(\beta)}, 
\label{EH}
\eea
where $\beta=1/T$ is a fictitious inverse temperature ("entanglement temperature") 
and $Z(\beta)=\tr e^{-\beta H_E}$ is a normalization constant chosen so that $\tr\rho_V=1$.
We will fix the normalization of $H_E$ later so that $\beta=2\pi$.

Let us now consider the case 
where this reduced density matrix of the subsystem $V$ 
has a conserved $U(1)$ charge ${Q_E}$.
The meaning of the conserved charge here is that the charge operator
${Q_E}$ defined on $V$ commutes 
with the entanglement Hamiltonian $H_E$: $[H_E,{Q_E}]=0$.
In this case, one can modify the reduced density matrix by introducing a chemical potential $\mu_E$
\bea
\tilde{\rho}^{\ }_V(\beta,\mu_E)=
{e^{-\beta H_E+\mue {Q_E}}\over Z(\beta,\mue)},
\label{charged reduced}
\eea
where 
$Z(\beta,\mue)=\tr e^{-\beta H_E+\mue {Q_E}}$ 
is a normalization constant to ensure 
${\tr_V} \tilde{\rho}_V=1$.
The entanglement Hamiltonian $H_E$ in (\ref{charged reduced}) is the same as $H_E$ in (\ref{EH})
and the state we are studying is the same pure state in this sense.

Since $\tilde{\rho}_V$ is a reduced density matrix, the charge operator
${Q_E}$ acts only on the Hilbert space of the subsystem $V$.
Therefore the chemical potential $\mue$ is different from the physical chemical potential
for which the conjugate charge operator acts on the entire system.
We call $\mue$ an entanglement chemical potential to distinguish it from the physical chemical potential.

By using this modified reduced density matrix, we define a charged \ren entropy by
\bea
S_{n}(\mue)
&=&
{1\over 1-n}\log \left[\left({e^{-\beta H_E}\over Z(\beta,\mue)}\right)^n e^{\mue Q}\right], 
\label{charged renyi def}
\eea
Note that this definition of the charged \ren entropy is slightly different from the one 
presented in Ref.\ \onlinecite{Belin:2013uta}.  
The chemical potential $\mue$ here is rescaled by $1/n$ than that in Ref.\ \onlinecite{Belin:2013uta},
which turns out to be more natural in topological theories discussed in this paper.

The charged entanglement entropy is defined by taking the $n\to 1$ limit
in the charged \ren entropy. It can be expressed as
\bea
S_{EE}(\mue)=-\sum_{i} p_i e^{\mue q_i} \log p_i,
\eea
where we introduced $p_i$ and $q_i$ as eigenvalues of $H_E$ and 
${Q_E}$:
\bea
{e^{-  \beta H_E} \over Z(\beta,\mue)} |i\rangle 
=p_i |i\rangle,~~ e^{\mue {Q_E}} |i\rangle = e^{\mue q_i} |i\rangle.
\label{charged EE spectrum}
\eea
Note that the factor $e^{\mue q_i}$ in (\ref{charged EE spectrum}) appears in the coefficient of the logarithm but not 
in the argument of it.
This asymmetry comes from the fact that the \ren parameter $n$ changes the effective entanglement temperature
from $\beta$ to $n\beta$ while keeping the the entanglement chemical potential $\mue$ intact as shown in the second line of
(\ref{charged renyi def}).
It is clear that this recovers the standard entanglement entropy in the limit $\mue=0$
\bea
S_{EE}(\mue=0)=-\sum_{i} p_i  \log p_i.
\eea

It is also useful to consider the modified reduced density matrix from the path integral point of view.
Let us consider a quantum field $\psi$. 
We denote the basis of the quantum field in the subsystem $V$ as $\psi_V$ or $\psi'_{V}$.
In the Euclidean path integral, the reduced density matrix of a ground state is described by
\begin{align}
\langle \psi_V |\rho_V| \psi^{'}_V \rangle &=
{1\over Z(\beta)}\langle \psi_V |e^{-\beta H_E}| \psi^{'}_V \rangle
\nonumber \\
& =
{1\over Z(\beta)}
\int 
\mathcal{D}\psi\,  e^{-S(\psi)}
\Big|_{\psi(t=0_-, \vec{x}\in V)=\psi_V\atop \psi(t=0_+, \vec{x}\in V)=\psi'_V},
\label{path-red}
\end{align}
where $t$ is a time coordinate and $\vec{x}$ is a spacial coordinate.
One can think that the entangling Hamiltonian 
$H_E$ is an evolution generator around the entangling surface $\pd V$, from
$(t=0_+, \vec{x}\in V)$ to
$(t=0_-, \vec{x}\in V)$,
and $\beta$ is an evolution time.

If we modify the entanglement Hamiltonian by introducing chemical
potential, 
as evolving from $t=0_+$ to $t=0_-$,
the state $ \psi'_V$ will be twisted,
where the amount of the twist 
depends on the total charge stored within the subsystem $V$ 
In path integral picture, this is achieved by turning on a
background $U(1)$ gauge field $A_{\mu}$ 
around the entangling surface: $A_{\theta}={ \mue \over 2\pi}$
where $\theta \in [0,2\pi]$ is the temporal angular coordinate 
around the entangling surface: 
\bea
\begin{split}
&\langle \psi_V |\tilde{\rho}_V| \psi^{'}_V \rangle\\
 =&{1\over Z(\beta)}\langle \psi_V |e^{-\beta H_E+\mue \int_{V} j^{\theta}}| \psi^{'}_V \rangle\\
 =&{1\over Z(\beta)}\int \mathcal{D}\psi\,
	e^{-S(\psi)+ {\mue\over 2\pi }\int_{0}^{2\pi}d\theta \int_{V}j^{\theta}}\Big|_{\psi(t=0_-, \vec{x}\in V)=\psi_V\atop \psi(t=0_+, \vec{x}\in V)=\psi'_V}.
\end{split}
\label{path-red-charged}
\eea
The gauge field holonomy around the entangling surface (the temporal angular circle) has non-zero value:
$\oint A_{\theta}d\theta=\mue$.
By Stokes' theorem, 
the effect of the background field is reduced to the insertion of 
a magnetic flux at the entangling surface: the field strength $F={\mue\over 2\pi} d(d\theta)$.

$\mue$ can be both real and imaginary.
When it is imaginary, it can be thought of as the Aharonov-Bohm phase 
as the state receives an additional phase as it goes around the
entangling surface. In this paper, we reserve $\mue$ as a real number and denote
$i\mue$ when we consider an imaginary entanglement chemical potential.
An imaginary chemical potential has been used in quantum Hall systems, for instance
\cite{Cappelli Imaginary}, and QCD, for instance \cite{RW Imaginary}.
One advantage of using the imaginary potential is that
 in the case where the conserved charge corresponds to angular momenta, 
 which we will consider in the next subsection, 
 a real chemical potential could lead to unwanted divergences in the partition function
 which come 
 from excitations moving faster than the speed of light at fixed angular potential
 in a non-compact space. 
 An imaginary chemical potential would allow us an extra flexibility 
 to extract useful information while fending off divergences.

It is also possible to turn on a background flux in directions other than the time direction.
For instance, 
if the entangling surface extends in a spacial direction $y$,
one can turn on a background gauge field $A_{y}\sim \phie$ in
the subsystem $V$.
In general, $A_y$ can be a function of space-time.
By repeating the same argument as above, one can  generalize the reduced density matrix as
\bea
\tilde{\rho}_V={e^{-\beta H_E + \phie J}\over Z(\beta,\phie)},
\eea
where $J$ is a current operator coupled with $A_y$.
We call $\phie$ an entanglement flux.


\subsection{Shifted Entanglement Entropies}

The charged entanglement entropy defined in the previous section assumes 
an internal $U(1)$ symmetry.
One can extend the idea to other $U(1)$ symmetries, 
such as a rotational or translational symmetry.
For simplicity, we consider a quantum field theory on a Euclidean flat space 
$(t,x_0,x_i)$ with $i\ge1$ and choose a half space $x_0>0$ for the subsystem $V$.
The metric is
\begin{align}
ds^2
&=dt^2+dx_0^2+dx_i^2
\nonumber \\
&=
r^2d\theta^2+dr^2+dx_i^2.
\end{align}
$(r,\theta)$ are polar coordinates of $(x_0,t)=(r\cos \theta, r\sin\theta)$.
From the expression (\ref{path-red}), the entanglement hamiltonian is a generator of the angular rotation in $(t,x_0)$ space and by choosing the normalization
so that $\beta=2\pi $, it is $H_E={i} {\pd \over  \pd\theta}$.

Observe that for the above choice,
the subsystem $V$ still has translational symmetry in $x_i$ directions.
I.e., the entanglement Hamiltonian $H_E= i {\pd \over \pd \theta}$ 
and the translation operator in $x_i$ direction $P^{i}={1\over i}{\pd\over \pd x_i}$ commute.
Hence, one can modify the reduced density matrix as
\bea
\tilde{\rho}_A={e^{-\beta H_E + a^{i} P_{i}}\over Z(\beta,a^i)},
\eea
where $a^i$ is a shift vector.
This quantity is closely related to a momentum polarization introduced in \cite{mom-pol}.

In addition to the flat entangling surface, 
one can also consider a spherical entangling surface where the subsystem $V$ is
inside a sphere. 
In this case, the momentum operator $P_i$ and the shift vector $a^i$ are replaced by the
angular momentum $J_i$ and the angular potential $\Omega^i$.
In the context of the gauge/gravity duality, 
the \ren entropy with the angular potential is dual to a rotating hyperbolic black hole
\cite{rotating ent} and it is called a rotating \ren entropy.
One difference between the momentum polarization and the rotating \ren entropy is that the shift vector breaks conformal symmetry whereas
the angular potential does not.

As in the case of the charged \ren entropy, when the shift vector (the angular potential) 
is turned on, the wave function is shifted around
the entangling surface. 
Therefore, one can think of this as a generalization of the Aharonov-Bohm effect.
It generates a dislocation along the entangling surface.
However, unlike the case of the charge $U(1)$, one cannot totally
localize the effect of the spacial $U(1)$ shift at the entangling surface.
This momentum polarization is similar to the entanglement chemical potential in the sense that it is generated by the time like component of the
energy momentum tensor
$P_i=\int_{V} T_{0i}$.
There is also a flux charge generated by spacial component of the energy momentum tensor
$\sigma_{ij}=\int_{V} T_{ij}$,
\bea
\tilde{\rho}_V={e^{-\beta H_E+ b^{ij}  \sigma_{ij}}\over Z(\beta, b^{ij})}
\eea
where $b^{ij}$ is an flux potential.

\section{The integer quantum Hall effect}

\subsection{The bulk Chern-Simons theory}
\label{bulk story}

As our first  example, let us consider a massive free fermion system in $d=2+1$ dimensions 
with non-zero Hall conductance (i.e., a Chern Insulator).
The entanglement charge $Q_E$ in this case is a particle number operator
restricted to $V$.
For free fermion problems, microscopic calculations of the (grand canonical) entanglement Hamiltonian 
and the charged entanglement entropy would be possible either analytical or by a mild use of computers. 
If we are interested only in the universal properties of the charged \ren and entanglement entropies, however,
it is more convenient to rewrite the problem
in terms of topological quantum field theories.
(This can be done, for example, by using the functional bosonization\cite{funcboso}.)

The effective Euclidean action for a Chern insulator with unit Chern number ($Ch=1$, say) is given as
\begin{align}
S={i \over 4\pi}\int a\wedge da 
+{{i \over 2\pi}}\int A^{ex}\wedge da.
\label{bf2}
\end{align}
Here, $A^{ex}$ is an external gauge potential,
and $a$ is the one form gauge field describing the electron current.

We are interested in turning on the entanglement chemical potential and the entanglement flux.
It is convenient to use the path integral formalism. In this case, the entanglement chemical potential is represented as a
gauge field around the entangling surface
and the entanglement flux is represented as a gauge field along the entangling surface.
To be more precise, let us consider a 2+1 dimensional flat space time $\mathbb{R}^3$ and compactify it by adding an infinity $\left\{\infty\right\}$.
The topology of the total space is a sphere $S^3$.
Let us introduce coordinates $(r, \theta, x)$.
\begin{align}
ds^2=\sin^2 {\pi r\over 2} d\theta^2+dr^2+\cos^2 {\pi r\over 2}  dx^2,
\end{align}
where $r$ is a distance coordinate from the entangling surface and takes values in $r\in[0,1]$;
$\theta$ is an angular coordinate around the entangling surface and takes values in $\theta\in[0,2\pi]$;
and finally
$x$ is an angular coordinate along the entangling surface and takes values in $x\in [0,2\pi]$.
At the entangling surface $r=0$, the $\theta$ circle shrinks to zero whereas $x$ circle is finite.
On the other hand, at $r=1$, the $x$ circle shrinks to zero whereas $\theta$ circle is finite.
One can see that this indeed makes a sphere.
In this coordinate, the entanglement chemical potential is $A^{ex}_{\theta}=\mue$  and the entanglement flux
is $A^{ex}_{x}=\phie $. Note that the normalization of the gauge potential $A^{ex}$ is different from the one in
(\ref{path-red-charged}).

From the second term of (\ref{bf2}), 
by doing partial integration, the entanglement chemical potential generates
a Wilson loop along the entangling surface
 \begin{align}
 \mue Q_E
 &={1\over 2\pi} \int A_{\theta}^{ex}d\theta\wedge da
 \nonumber\\
 &={\mue} \int_V\pd_{[r}a_{x]}drdx
 \nonumber \\
 &=-{\mue} \oint_{r=0}a_xdx,
\label{Wilchem}
 \end{align}
whereas the entanglement flux generates a Wilson loop 
 \begin{align}
 \phie J=&{1\over 2\pi}\int A_{x}^{ex}dx \wedge da
	 \nonumber \\
 =&\phie \int_V\pd_{[r}a_{\theta]}drd\theta
	 \nonumber \\
 =&\phie \oint_{r=1}a_{\theta} d\theta.
\label{Wilflux}
\end{align}
Notice that although $r$ has two boundaries $r=0$ and $r=1$,
$a_{x(\theta)}$ must vanish at $r=1(0)$ from regularity.
With these Wilson loops, 
the action is now given by
\begin{align}
S&={i \over 4\pi}\int a\wedge da
+\mue\oint_{r=0} a_{x}dx -\phie \oint_{r=1}a_{\theta}d\theta.
\label{CS CI-1}
\end{align}
The two Wilson loops link each other. 
Therefore, evaluating the partition function gives 
\bea
Z(\mue,\phie)=\exp\left({2\pi i } \mue\phie  Lk \right),
\label{link wilson}
\eea
where $Lk$ is the Gauss linking number
\bea
Lk&=&{1\over 4\pi}\int_{r=0}dx^{\mu} \int_{r=1}dy^{\nu}\epsilon_{\mu\nu\rho}{(x-y)^{\rho}\over |x-y|^3} \cr
&=&1.
\eea


The charged \ren entropy, 
\begin{align}
S_{n}(\mue,\phie)=
{1\over 1-n}\log \left[\left({e^{-\beta H_E}\over Z(\beta,\mue,\phie)}\right)^n e^{\mue Q+\phie J}\right], 
\label{Sn}
\end{align}
can be computed from $Z(n\beta,\mue,\phie)$,
for which the universal part does not depend on $n$.
Therefore the \ren entropy is
\bea
S_{n}(\mue,\phie)=\log Z(\beta,\mue,\phie).
\eea
From (\ref{link wilson}), the charged entanglement entropy is
\bea
\gamma_{c}(\mue,\phie)= 2\pi i \mue\phie .
\eea

When $|Ch|>1$, 
the hydrodynamic Chern-Simons theory for Chern insulators 
is given in terms of 
two gauge fields $a$ (``the statistical gauge field") and 
$b$ (``the hydrodynamic gauge field"), and described by the action 
\begin{align}
	S=\frac{i Ch}{4\pi}\int a\wedge da
	-
	\frac{i}{2\pi} \int b\wedge (da-dA^{ex}).
\end{align}
Introducing a two-component notation $(a^1,a^2):= (a,b)$, 
the hydrodynamic theory can be also written as
\begin{align}
	S=\frac{iK_{IJ}}{4\pi} \int a^I\wedge da^J,
	\quad
I,J=1,2, 
\end{align}
where the $K$-matrix is given by
\begin{align}
	K = \left(
		\begin{array}{cc}
			Ch & -1 \\
			-1 & 0
		\end{array}
	\right).
\end{align}
By integrating over both $a$ and $b$,
the Chern-Simons term
\begin{align}
	\frac{-i Ch}{4\pi} \int A^{ex}\wedge dA^{ex},
\end{align}
is generated for the effective action of the external field,
indicating that the Hall conductivity of the system is quantized. 
Observe also that the determinant of the $K$ matrix is
$|\mathrm{det}\, K|=1$ as it should be since Chern insulators 
are not topologically ordered. 
By introducing the entanglement chemical potential and flux, 
let us consider
\begin{align}
	S&= \frac{iCh}{4\pi} \int a \wedge da-
	\frac{i}{2\pi} \int b \wedge da
	\nonumber \\
	&\quad
	-\mu_E \oint b_xdx + \phi_E \oint b_{\theta}d\theta,
\end{align}
where we noted
the current 
is given by
\begin{align}
j^{\mu}=\delta S/\delta A^{ex}_{\mu} 
= \frac{1}{2\pi} \epsilon^{\mu\nu\lambda} \partial_{\nu} b_{\lambda}. 
\end{align}
From this,
one sees that the $\log Z$ picks up the contribution
\begin{align}
	\gamma_c
	=
	\mu_E \phi_E t_I t_J \times 2\pi i K^{-1}_{IJ}
\end{align}
where 
\begin{align}
	t= \left(
		\begin{array}{c}
			0 \\
			1
		\end{array}
	\right),
	\quad
	K^{-1}=
	\left(
		\begin{array}{cc}
			0 & 1 \\
			1 & -Ch
	\end{array}
	\right). 
\end{align}
Thus, the charged topological entanglement entropy is given by
\begin{align}
	\gamma_c = Ch\times 2\pi i \mu_E \phi_E.  
\end{align}


\subsection{Boundary theory}
\label{sec CFT CI}

It is known that 
the relevant part of the reduced density matrix of topological phases is described by the edge theories
\cite{QKL}.
Therefore one can reproduce the
the charged entanglement entropy obtained in the previous section from the edge theory point of view.

\subsubsection{Boson description}

We are interested in a Chern insulator with $Ch=1$.
It is known that the low energy modes of the entanglement Hamiltonian
essentially the same as the physical edge Hamiltonian\cite{Ryu-Hatsugai,QKL,LiHaldane}.
Hence, 
we consider a chiral free boson $\Phi$ living on the entangling surface.
The chiral boson satisfies the {constraint} 
\bea
(\pd_{t}-\pd_{x})\Phi=0,
\label{eomboson}
\eea
where $x$ is the spatial coordinate along the entangling surface.
We compactify it into $2\pi$.
The Hamiltonian is
\bea
{H_E}={1\over 4\pi}\int dx (\pd_x \Phi)^2.
\label{Hboson}
\eea
The entanglement chemical potential and the entanglement flux act as the chemical potential and
the Berry gauge potential along the spacial direction respectively.
The charge term induced by the chemical potential is
\bea
{Q_E}=\int dx \pd_x \Phi.
\eea
The entanglement flux changes the spatial boundary condition for
$\Phi$,
\bea
\Phi(x+2\pi,t)=\Phi(x,t)+2\pi n+2\pi \phie,
\label{bcboson}
\eea
where $n$ is an integer.

From (\ref{eomboson}) and (\ref{bcboson}),
the mode expansion of $\Phi$ is
\begin{align}
\Phi(t,x)=\Phi_0+ p_L (t+x)+i\sum_{m\in\mathbb{Z}\atop m\neq 0}{a_{m}\over m}e^{-im(t+x)},
\end{align}
where $p_{L}=(n+\phie)$ is the momentum and $a_{m}$ is a creation (annihilation) operator for $m < (>)0$.

We would like to compute a partition function of the boundary 
theory
on a spacetime torus whose modular parameter is $\tau=\tau_1+i\tau_2$.
In our case, the spacial shift is not important so we set $\tau_1=0$.
Therefore, the partition function is 
\begin{align}
Z(\tau,\mue)
&=\mathrm{Tr}\, \exp\left(-\tau_2 {H_E}+i \mue {Q_E}\right)
\nonumber \\
\quad 
&=\eta(\tau)^{-1}\sum_{n\in \mathbb{Z}}
e^{
-\tau_2 (n+\phie)^2+2\pi i \mue (n+\phie)
},
\label{integerpartition}
\end{align}
where 
\begin{align}
\eta(\tau)&=q^{1\over 24}\prod_{n}(1-q^{n}),
\end{align}
is the Dedekind eta function.

We are interested in the thermodynamical limit $\tau_2\to 0$ of this partition function. To evaluate the partition function in this limit,
it is convenient to perform the Poisson resummation:
\begin{equation}
\begin{split}
&\quad Z(\tau,\mue)\\
 &=
 \eta(\tau)^{-1}
 \sum_{\tilde{n}\in \mathbb{Z}} 
 \int_{-\infty}^{\infty} d{n}
e^{-\tau_2 (n+\phie)^2
 +2\pi i \mue (n+\phie) +2\pi i n\tilde{n}}
 \nonumber \\
&=\sqrt{1\over \tau}\eta(\tau)^{-1}\sum_{\tilde{n}\in \mathbb{Z}}
e^{-{\pi^2(\mue-\tilde{n})^2\over \tau_2} +2\pi i \tilde{n}\phie }.
\label{part func CI}
\end{split}
\end{equation}
In the thermodynamical limit $\tau_2\to 0$, one can see that
only $\tilde{n}$ which minimizes $(\mue-\tilde{n})^2$
contributes to the partition function.
As we saw in the bulk computation, 
the chemical potential $\mue$ is the coefficient of the gauge field holonomy in the Wilson loop. 
If we require that this Wilson loop is a dynamical operator 
in the theory in the sense that a state created by acting the Wilson loop on a physical state
is still a physical state, then the coefficient has to take
discrete values consistent with large gauge transformations. 
On the other hand, if such Wilson loop is treated as an external source, 
the values of $\mue$ and $\phi$ are arbitrary.
In this case, the partition function (\ref{part func CI}) decays to zero as $\tau\to0$.
On the other hand, if we choose the values of $\mue$ and $\phi$ 
so that they are physical states of the theory, there must be $\tilde{n}$ satisfying $\tilde{n}=\mue$. 
Then the partition function becomes
\begin{align}
\log Z(\tau,\mue)\Big|_{\tau_2\to0}=\log\left({1\over \eta(\tau)} \frac{1}{\sqrt{\tau}}\right)+2\pi i \mue\phie.
\label{entropy001}
\end{align}
The first term of the right hand side gives the area law term,
which is non-universal and divergent.
Note also that it depends neither on $\phie$ or $\mue$.
On the other hand,
the second term is proportional to $2\pi i \mue \phie$,
whose coefficient is universal (simply one in this case);
we call the second term the charged topological entanglement entropy,
which is given by 
\bea
\gamma_c= 2\pi i \mue\phie.
\label{ctee}
\eea
In the above we considered the case of $Ch=1$. 
For a general Ch, 
which corresponds to Ch chiral edge states, we will have
\bea
\gamma_c={Ch} \times ( 2\pi i \mue\phie).
\label{ctee2}
\eea
As can be seen in (\ref{bcboson}),
the role of the entanglement flux $\phie$ is to shift the modes,
while the entanglement chemical potential $\mue$ detects the change of the charge, or the modes
(\ref{integerpartition}). Therefore, the charged topological entanglement entropy (\ref{ctee2}) is proportional to the product of $\phie$ and $\mue$,
and the physical information is the coefficient of it: $Ch$.

\subsubsection{Fermion description}

Alternatively, 
one can understand the above computation in terms of a free fermion system.
Let us consider a fermion $\psi(x,t)$ with the Hamiltonian 
\bea
{H_E}={i\over 2\pi}\int dx \psi^{\dagger}\pd_{x}\psi.
\eea
The entanglement chemical potential and the entanglement flux change 
the temporal and spatial boundary conditions as
\begin{align}
\psi(x+2\pi,t)&=-e^{2\pi i\phie}\psi(x,t),
\nonumber \\
\psi(x,t+2\pi)&=-e^{2\pi i\mue}\psi(x,t).
\end{align}
The consistent mode expansion is
\begin{align}
&
\psi(x,t)=\sum_{r\in \mathbb{Z}+\phi_E-{1\over 2}} a_{r}e^{-ir(x+t)},
\nonumber \\
&
\mbox{with}
\quad 
\{a_r,a^{\dagger}_s\}=\delta_{r,s}.
\end{align}
For a given $\phi$, the vacuum is defined by
\bea
a_{n+\phie-{1\over 2}}|0\rangle_{\phie}=0 ~~~\text{for}~~n=1,2,\cdots .
\label{fermion-vac}
\eea
By inserting the mode expansion into the Hamiltonian, and 
normal order with respect to the vacuum,  
one obtains
\begin{align}
	{H_E}&=
\sum_{n=-\infty}^{\infty}\left(n+\phie-{1\over 2} \right) : a_{n+\phie-{1\over 2}}^{\dagger}  a_{n+\phie-{1\over 2}}:
\nonumber \\
&\quad 
-{1\over 24}+{\phie^2\over 2}.
\end{align}
The chemical potential detects the number of fermions, so it is described by the insertion of operator $g$ that satisfies
$gag^{-1}=e^{-2\pi i\mue}a$
Notice that 
the vacuum (\ref{fermion-vac}) itself has non-zero fermion number $g|0\rangle_{\phie}=e^{2\pi i \mue \phie}|0\rangle_{\phie}$.
Taking this into account,
the partition function is \cite{AlvarezGaume:1986es}
\begin{align}
\mathrm{Det}\, (\phie,\mue)
&=\Tr gq^{H}
\nonumber \\
&=
e^{2\pi i \left[\mue\phie+\tau \left(-{1\over 24}+{\phie^2\over 2}\right)\right]}
\nonumber \\
&\quad 
\times 
\prod_{n=1}^{\infty}(1+e^{2\pi i\tau(n+\phie-{1\over 2})+2\pi i \mue}) 
\nonumber \\
&\quad
\times(1+e^{2\pi i\tau (n-\phie-{1\over 2})-2\pi i \mue}). 
\label{fermi-partition}
\end{align}
This fermion partition function is indeed equivalent to the bosonic one in the previous section due to the following formula
\bea
\mathrm{Det}\, (\phi,\theta)=
\frac{1}{\eta(\tau)}
\vartheta
\begin{bmatrix}
\phi \\
\theta
\end{bmatrix}
(0|\tau),
\eea
 where
\begin{align}
{\vartheta}
\begin{bmatrix}
\phi \\
\theta
\end{bmatrix}
(0|\tau)
&=\sum_{n}e^{i\pi (n+\phie)^2\tau+2\pi i (n+\phie)\theta}.
\label{fermion par function}
\end{align}
By taking the thermodynamical limit $\tau_2\to0$, one obtains the charged topological entanglement entropy (\ref{ctee}).


\section{Symmetry Protected Topological Phases}
\label{SPT story}

In the previous section, we consider a Chern insulator which has chiral edge modes.
In this section, we consider a general class of SPT phases
in (2+1) dimensions which have a Chern-Simons theory description \cite{Ashvin}.
The potential edge modes of SPT phases are non-chiral; there are same numbers of right moving and left moving
edge modes. Those phases are unstable under perturbation and the edge modes are gapped out unless we impose some symmetries.

\subsection{Bosonic  SPT phases}

\subsubsection{Bulk and boundary field theory descriptions}

A convenient description of a wide class of (2+1)-dimensional SPT phases 
is given in terms of the multi-component $K$-matrix Chern-Simons theories and 
their edge theories. 
\cite{Ashvin}
Let us consider a Chern-Simons theory with a characteristic $K$-matrix 
and a charge vector $t^I$ given in terms of the Euclidean action, 
\begin{align}
 S&={i\over 4\pi}\int K_{IJ}a^{I}\wedge da^{J}
 + \int \frac{i}{2\pi} t^I A^{ex}  \wedge da^I,
 \label{charged CS}
 \end{align}
where $I,J$ runs from $1$ to $M$, and $K_{IJ}$ is a symmetric matrix taking integral
values.
For SPT phases, 
there is
no
 ground state degeneracy 
and $K_{IJ}$ has unit determinant: $|\det K|=1$.
Note that $K$ matrices connected by $GL(N,\mathbb{Z})$ transformations are physically equivalent.

In this paper, we consider only $2\times 2$ $K$ matrices for SPT phases. This is sufficient in many cases
since
more general cases can be understood
as a direct sum of the $2\times 2$ $K$ matrices.
For a bosonic system with $\det K=-1$, one can always
 bring the $K$-matrix into a canonical form, 
$K = \sigma_x$, by using $GL(2,\mathbb{Z})$ transformation,
where $\sigma_x$ is the Pauli-matrix,
$
\sigma_x=
\begin{pmatrix}
0 & 1 \\
1 & 0
\end{pmatrix}
$.

The coupling to an external gauge field $A^{ex}$ is specified by the $M$-component charge vector $t^I$.
In principle, more than one type of external gauge fields can be introduced, but for our purposes below,
it suffices to consider a single external $U(1)$ gauge field. 
In this section,
we will consider SPT phases protected by $\mathbb{Z}_N$ symmetry, which can be ''embed`` into a global $U(1)$ symmetry. 
I.e., $\mathbb{Z}_N$ symmetry can be thought of as obtained by spontaneously breaking the $U(1)$ symmetry.
(See below.)

It is also convenient to look at the corresponding edge theory,
which consists of two components of scalars $\phi^{I=1,2}$. 
They are compactified by $2\pi$, $\phi^I(x) = \phi^I(x) + {2\pi}$.
An $\mathbb{Z}_N\times \mathbb{Z}_N$ 
symmetry transformation can act on the scalars as 
\be
\phi^I(x) \to \phi^I(x) + {2\pi \over N} L^I,
\label{transform}
\ee
where $L=(k, l)^T\in \mathbb{Z}_N\times \mathbb{Z}_N$ is an integer vector.
We will focus on $\mathbb{Z}_N$ symmetry within  $\mathbb{Z}_N\times \mathbb{Z}_N$
by imposing $l=qk$ with $q\in \mathbb{Z}_N$.
By using $GL(2,\mathbb{Z})$ transformation, one can always take it into
$(k,l)=(1, q)$.

The transformation law \eqref{transform}
can be deduced from the following generic argument.
To implement the symmetry $\mathbb{Z}_N$ in the $K$-matrix theory, 
we need to specify how the $K$ matrix and the edge modes $\phi^I$ transform under $g$,
which is the generator of $\mathbb{Z}_N$ and satisfies $g^N=e$ (identity).
Since it is a symmetry, we require its action on the $K$-matrix, denoted by $W^g$,
leave the $K$ matrix unchanged, 
\bea
K&=&(W^g)^{T}KW^g,
\label{cond 1}
\eea
where $W^g\in GL(2,\mathbb{Z})$.
The symmetry $g$ acts on the edge modes as
 \bea
 \phi_{I}&\to& \sum_{J} W^{g}_{IJ} \phi_{J}+\delta \phi_I^g,
 \label{sym trans edge}
 \eea
where $\delta \phi^g_{I}$ is a constant phase rotation.
$W_{IJ}^{g}$ and $\delta \phi_I^g$ are determined 
from the requirement $g^N=e$,
which
gives 
\begin{align}
&(W_{IJ})^N=\delta_{IJ}, \cr
&\sum_{a=1}^{N}(W_{IJ})^{a-1}\delta \phi^g =
\begin{pmatrix}
0 \\
0
\end{pmatrix}
~~ mod~~2\pi.
\end{align}
Combining it with the constraint (\ref{cond 1})
we obtain
\begin{align}
W_{IJ}=\delta_{IJ},
\quad 
\delta \phi^g={2\pi k\over N}
\begin{pmatrix}
1 \\
q
\end{pmatrix},
\label{g-tranform}
\end{align}
where $k$ is an integer whose greatest common divisor with $N$ is 1.

In terms of the edge theory, the spontaneous symmetry breaking of $U(1)$ down to $\mathbb{Z}_N$
can be described as follows. 
A bulk quasiparticle excitation characterized by an integer vector $l_I$ is mapped to
a boundary field $e^{il_{I}\phi_{I}}$,
which can be added to the action of the edge theory. 
In general such Higgs terms can be written as
\bea
S=\sum_{l}\int dxdt\, C_{l}\cos(l_{I}\phi_{I}+\al_{l}),
\label{higgs}
\eea
where $C_{l}$ and $\al_{l}$ are constants.
Once we require $\mathbb{Z}_N$, 
we consider only those integer vectors $l$ for which
the corresponding cosine term is invariant
invariant under the $g$ transformation $\delta \phi^g$ in (\ref{g-tranform}).
If there is a set of such terms allowed by symmetry, and if they can localize all the degrees of freedom
(i.e., if they can completely gap out the edge), 
then it is a trivial phase.
On the other hand, 
there is no cosine term that can gap out the edge, 
we have a SPT phase protected by $\mathbb{Z}_N$ symmetry.
As shown in  \cite{Ashvin}
$q=0$ is the trivial phase while $q\neq 0$, $mod~N$ are non-trivial phases.

In the following, 
we will show that the charged entanglement entropy can capture this classification of 
SPT phases protected by $\mathbb{Z}_N$ symmetry.

\subsubsection{Charged entanglement entropy from the bulk field theory}

To turn on the entanglement chemical potential and the entanglement flux, 
we set $A^{ex}_{\theta}=\mue$,
$A^{ex}_{x}=\phie$,
and
$\bf{t}=(k,l)^T$, where $k,l\in \mathbb{Z}_N$.
The coupling to the external background fields generates Wilson loops:
 \begin{align}
 \mue Q_E
 &={1\over 2\pi} \int t^{I}A_{\theta}^{ex}d\theta\wedge da^{I}
 \nonumber\\
 &={\mue} {t}^{I} \int_V\pd_{[r}a^{I}_{x]}drdx
 \nonumber \\
 &=-{\mue}{t}^{I} \oint_{r=0}dx a^{I}_{x},
\label{SPTcharge}
 \end{align}
 and
 \begin{align}
 \phie J=&{1\over 2\pi}\int t^{I}A_{x}^{ex}dx \wedge da^{I}
	 \nonumber \\
 =&{\phie}{t}^{I}\int_V\pd_{[r}a^{I}_{\theta]}drd\theta
	 \nonumber \\
 =&{\phie}{t}^{I}
\oint_{r=1}d\theta a^{I}_{\theta}.
\label{SPTflux}
\end{align}
From the action (\ref{charged CS}), 
the gauge field propagators 
(in the covariant Lorentz gauge $\delta^{\mu\nu}\pd_{\mu}a_{\nu}=0$)  are
\bea
\left\langle a_{\mu}^I(x)a_{\nu}^J(y) \right\rangle={iK^{-1}_{IJ} \over 2} \epsilon_{\mu\nu\alpha}{(x-y)^{\alpha}\over |x-y|^3}.
\label{a-prop}
\eea
By using this, one can compute the link number of the two Wilson loops
\bea
\left\langle \oint_{r=0}dx^{\mu}a^{I}_{\mu}(x) 
\oint_{r=1}dx^{\nu}{a}^{J}_{\nu}(y)\right\rangle=2\pi i K^{-1}_{IJ}.
\eea
Then the partition function picks up a phase factor
\begin{align}
&
\left({\mue} {\phie} \right)
\left({t}_{I} {t}_{J}\right)
	\left\langle
\oint_{r=0}dy a^{I}_{y}
\oint_{r=1}d\theta a^{J}_{\theta} \right\rangle
\nonumber \\
&=({t}_{I} K^{-1}_{IJ}{t}_{J})
\left(2\pi i{\mue}{\phie}\right)
\nonumber \\
&=\left(2 k l\right)
\left(2\pi i{\mue}{\phie}\right).
\label{CSpartition}
\end{align}
$\phie$ is quantized in the unit of $1/N$ because of $\mathbb{Z}_N$ symmetry.
For example, consider the familiar case of superconductors, 
where the flux $2\pi \phie=0,\pi$ mod $2\pi$
because of $\mathbb{Z}_2$ symmetry.
Similar quantization is required for the chemical potential $\mue$ once $\mathbb{Z}_N$ symmetry is gauged.
The classification of the SPT phases, however, does not explicitly depend on the concrete values of $\phie$ and $\mue$.
Therefore we do not specify them here.
By setting $k=1,$  $l=q$, Eq.\ (\ref{CSpartition}) is rewritten as
\bea
\gamma_{c}=2q(2\pi i \mue\phie).
\label{CSpartition2}
\eea
The topological part of the charged entanglement entropy is  $\mathscr{C}={2q }$.
The trivial phase is $q=0~mod~N$, and indeed the charged topological entanglement entropy vanishes in this case.

One can see that in the case of $\mathbb{Z}_N$ symmetry, 
the charged entanglement entropy is essentially equivalent to the
braiding statistical phase of vortices (quasiparticles). 
The classification of SPT phases based on the braiding statistics in 2d systems is considered in
Ref.\ \onlinecite{Cheng Gu}.
When the vortices characterized by $\mathbb{Z}_N$ charge vector ${t}_I$ are exchanged, they
generate a phase $-\pi {t}_I^{T}K_{IJ}^{-1}{t}_J/N^2$. Since the braiding of a unit charge vortex generates
a phase $2\pi/N$, ${t}_I^{T}K_{IJ}^{-1}{t}_{J}$ is determined modulo $2N$.
The charged topological entanglement entropy $\mathscr{C}$ is ${t}_{I}^{T}K_{IJ}^{-1}{t}_{J}=2q$ in this model.
The distinct phases are $q\in [0,N-1]$ and
therefore in the bosonic case, the classification  is $\mathbb{Z}_N$.

\subsubsection{Charged entanglement entropy from the edge theory}

 We can also study the charged entanglement entropy from the edge CFT point of view.
Following \cite{wenx, LiHaldane, QKL}, we consider the edge hamiltonian as an entanglement hamiltonian.
In addition to the edge hamiltonian,
there are terms coming from the coupling to the external fields.
From (\ref{SPTcharge}) and (\ref{SPTflux}) and the bulk-boundary relation
$a^{I}_{i}=\pd_{i} \phi^{I}$, the boundary CFT partition function is
$Z(\tau)=\tr \exp\left(-{\tau_2} H_E +i \mue Q_E\right)$
where
\begin{align}
&
\tau_2 H_E-i\mue Q_E=
	{{\tau_2} \over 2\pi}\int_{0}^{2\pi} dx {1\over 2}\left[(\pd_x \phi_{L})^2+(\pd_x \phi_{R})^2\right]
\nonumber \\
&\quad -i\mue \int_{0}^{2\pi} dx 
\left(k_2\pd_{x}\phi^2+l_2\pd_{x}\phi^1\right).
 \label{boundary-hamil}
 \end{align}
Here we introduced the subscript 2 for the components of the charge vector $(k_2,l_2)$.
 This subscript is for the spatial boundary condition, and we will use the subscript 1, $(k_1,l_1)$, for the temporal boundary
 condition. This is simply for a technical reason and we will remove them and set  $(k_1,l_1)=(k_2,l_2)=(k,l)$ at the end of the computation.
The left-moving and the right-moving modes $\phi_{L}$ and $\phi_R$ are defined by combining $\phi^{1}$ and $\phi^2$ as
\bea
\phi_L&=&\sqrt{{1\over 2r}}(\phi^1+r \phi^2),\cr
\phi_R&=&\sqrt{{1\over 2r}}(\phi^1-r \phi^2),
\eea
in which $r$ is the compactification radius (the Luttinger parameter) and depends on 
the microscopic details of the edge Hamiltonian.
We fix the length of the spacial direction to $2\pi$.
The effect of the entanglement flux is encoded in the quantization of the modes.
Each mode $\phi_1$ and $\phi_2$ has the following boundary condition along the spacial direction
\begin{align}
\phi_1(x+2\pi,t)&=\phi_1(x,t)+2\pi (n+{k_1\over N}),\cr
\phi_2(x+2\pi,t)&=\phi_2(x,t)+2\pi (m+{l_1\over N}),
\label{ident}
\end{align}
where the shifts $k_1/N, l_1/N$ are due to the flux $\phie$.
We emphasize that the flux is discretized as we saw in the bulk computation.
We see that the entanglement chemical potential $\mue$ generates the twist boundary conditions in the time direction
whereas the entanglement flux generates the twist boundary conditions in the spacial direction.


From the equations of motion, the mode expansion for $\phi_L$ and $\phi_R$ are
\begin{align}
\phi_L(x,t)&=\phi_{L,0}+p_L(t+x)+i\sum_{n\neq 0}{a_n\over n}e^{-in(t+x)},
\cr
\phi_R(x,t)&=\phi_{R,0}+p_R(t-x)+i\sum_{n\neq 0}{b_n\over n}e^{-in(t-x)}.
\end{align}
The identifications (\ref{ident}) determines the momenta as
\begin{align}
 p_{L}&=\sqrt{{1\over 2r}}\left[(n+{k_1\over N})+r(m+{l_1\over N})\right],\cr
 p_{R}&=\sqrt{{1\over 2r}}\left[-(n+{k_1\over N})+r(m+{l_1\over N})\right].
 \end{align}
By using the mode expansion, the modified hamiltonian is written as
\begin{align}
&\quad 
\tau_2H_E-i\mue Q_E
\nonumber \\
&=\tau_2
\left[{p^2_L\over 2}+{p^2_{R}\over 2}+(\sum a_{-n}a_{n}-{1\over 12}+b_{-n}b_{n}-{1\over 12})
\right]
\nonumber \\
&\quad 
-i\mue
\left[k_2\sqrt{{1\over 2r}}(p_L+p_R)+l_2\sqrt{{r\over 2}}(p_L-p_R)\right].
\end{align}

The universal part of the entanglement entropy can be obtained by taking the thermodynamical limit $\tau_2\to0$ or equivalently
taking the large area limit $L/\beta\to \infty$ where $L$ is the size of the subsystem $V$ \cite{wenx}.
$\tau_2\to0$ is a high temperature limit and all the modes contribute to the partition function.
It is more convenient to take the modular transformation $\tau\to -1/\tau$ and compute the partition function of
the low energy limit.  In this case, only the ground state contributes to the partition function.


To this end, we rewrite the partition function so that the modular transformation of the partition function
gives the same partition function up to the phase factor and re-parametrization.
Following the method used in \cite{Shinsei-GLaugh},
we first perform Poisson re-summation for $n$
 \bea
 \int dn \exp(-(\tau_2 {H_E}-i\mue {Q_E})) \exp(-2\pi i n \tilde{n}),
 \eea
 where $\tilde{n}\in \mathbb{Z}$.
After redefining $\tilde{n}$ to $n$, it is
\begin{align}
&Z(\tau)_{k_1,k_2,l_1,l_2}
= {1\over \eta(\tau)}\sqrt{{r\over \tau_2}}
\sum_{n,m=-\infty}^{\infty}
\nonumber \\
&\quad 
\times
e^{
-{r\pi\over \tau_2}(\mue l_2 -n)^2+2\pi i n {k_1\over N}
 -\pi \tau_2 r (m+{l_1\over N})^2+2\pi i \mue k_2(m+{l_1\over N})}.
 \end{align}
One can show that under the $S$ transformation  ($\tau_2\to {1\over \tau_2}$) and the following reparameterization
\begin{align}
&
m\to n, \quad 
n\to -m, \quad 
l_1\to -\mue N l_2, \quad 
l_2 \to {l_1\over \mue N}, 
\nonumber \\
&
k_1\to -\mue N k_2, \quad 
k_2\to {k_1\over \mue N}.
\end{align}
the partition function transforms as
\begin{align}
Z(\tau)\to Z(-1/\tau)\exp\Big(2\pi i{\mue\over N}(k_1l_2+k_2l_1)\Big).
 \label{CFTpartition}
\end{align}
 So far we set $n,m\in\mathbb{Z}$. 
 However, as we saw in the example of the Chern insulator (Sec.\ \ref{sec CFT CI}),
  in order for the fractional flux and chemical potential be inside the
 Hilbert space, we have to gauge the $\mathbb{Z}_N\times \mathbb{Z}_N$ symmetry. So $n$ and $m$ should take values in
 $\mathbb{Z}_N$.
 Then
 \bea
 Z(\tau_2)_{-k_2,k_1,-l_2,l_1}\Big|_{\tau_2\to \infty}={1\over \eta(\tau)}\sqrt{{r\over \tau_2}},
\eea
and
\bea
\begin{split}
&\log  Z(1/\tau_2)_{k_1,k_2,l_1,l_2}\Big|_{\tau_2\to\infty}\\
=& \log {1\over \eta(\tau)}\sqrt{{r\over \tau_2}}+
\Big(2\pi i{\mue\over N}(k_1l_2+k_2l_1)\Big).
 \end{split}
 \label{bosonSPTedge}
\eea
By setting $\mue=1/N$, 
one can reproduce the bulk result (\ref{CSpartition}).

 To obtain the charged entanglement entropy, we use the thermodynamical relation, 
 $S=-{\pd\over \pd T} T\log Z $. 
 In our case the inverse temperature $1/T$ is the \ren parameter
 $n$ which is multiplied by both $H_E$ and $Q_E$, rather than $\tau_2$ which acts only on $H_E$.
By setting $k_1=k_2=1$ and $l_1=l_2=q$
 one can show that the universal part of the charged entanglement entropy $S_{EE}(\mue,\phie)$ is
 \bea
 \gamma_c= ({2q})(2\pi i \mue \phie),
 \label{charged TEE CFT}
 \eea
which agrees with (\ref{CSpartition2}).

  The construction here is along the same vain as gauging the global symmetry by coupling it to a background field, while introducing a symmetry twist that leads to non-trivial winding number. The winding number we have calculated above can be recovered in discrete lattice models that realizes 
  SPT phases, such as the Dijkgraaf-Witten type models. See for example \cite{Hungwen}.

The topological entanglement entropy in topologically ordered system measures
the quantum dimension of a system which can be computed from the $S$ modular transformation of the partition function.
As seen from the above calculations, 
the charged entanglement entropy is related to the $S$ modular transformation as well. 
The difference is that the topological entanglement entropy is a real number,
while the universal part of the charged entanglement entropy is a phase.
In this sense, it can be interpreted as a diagnosis of anomaly.
In \cite{Shinsei-GLaugh}, the unremovable phase factors of the partition function under the modular transformations
is used as a diagnosis of the global gravitational anomaly in gauged SPT phases.
When the global symmetry of SPT phases is gauged, the system becomes a topologically ordered system \cite{GuLev}.
Although our computations are similar to \cite{Shinsei-GLaugh}, 
we do not gauge the symmetry to compute the charged entanglement entropy.

As a final comment, our method can be straightforwardly generalized to other SPT phases with on-site symmetries.
One interesting example is the bosonic SPT phases with $U(1)$ symmetry, 
which are classified by an integer;
Different phases can be distinguished by their quantized Hall conductance $\sigma=2q$, with $q\in \mathbb{Z}$.
The calculation of charged topological entanglement entropy is similar to the $\mathbb{Z}_N$ symmetry case. For simplicity,
here we briefly show the edge theory calculation.
Under the $U(1)$ symmetry transformation, the edge modes $\phi^I$ change as
Eq.\ (\ref{transform})
\be
\phi^I(x) \to \phi^I(x) + \phi_E L^I,
\ee
where $\phi_E$ is a continuous variable instead of a discrete variable in the 
{$\mathbb{Z}_N$} symmetry case, and $L^T=(1,q)$ with $q\in \mathbb{Z}$.
I.e., 
\bea
W=I, \quad 
\delta \phi=\theta {\bf t},
\quad
{\bf t}=\left(
\begin{array}{c}
	1 \\
	q
\end{array}
\right). 
\eea
The boundary condition along the spacial direction reads
\begin{align}
	\phi_1(x+2\pi,t)&=\phi_1(x,t)+2\pi (n+\phi_E),
	\nonumber \\
	\phi_2(x+2\pi,t)&=\phi_2(x,t)+2\pi (m+q\phi_E).
\end{align}
Following the procedures in the $\mathbb{Z}_N$ symmetry case, one can find that the charged topological entanglement entropy has the following expression
\begin{equation}
\gamma_c=(2q)(2\pi i \mu_E\phi_E),
\label{gammaBosonU1}
\end{equation}
i.e., the topological number $ \mathscr{C}$ in Eq.\ (\ref{ctee}) is nothing but the Hall conductance $2q$,
which parallels with the story in Sec. \ref{bulk story} and Sec. \ref{sec CFT CI}.

\subsection{Fermionic SPT phases}

\subsubsection{Bulk and boundary field theory descriptions}
One can extend the above analysis for bosonic SPT phases to fermionic SPT phases in (2+1) dimensions.
by considering the $K$-matrix theory description 
(both in the bulk and at the edge) of fermionic SPT phases.
The simplest case is when the $K$-matrix is $\sigma_z$.

As in the bosonic case, 
to diagnose the existence of gapless edge modes, 
one studies whether
there are bosonic degrees of freedom that take expectation value and gap out the edge modes without
breaking symmetries.
Since fermions cannot condense, a condensate must come from degrees of freedom consisting of
even number of fermions.
Therefore, identity operation in fermionic systems is not just a trivial transformations
but also changes the sign of all fermions.
This makes the fermionic symmetry projective.
The corresponding symmetry transformation is
\begin{align}
W^{g}_{IJ}=\delta_{IJ}, 
	\quad 
\delta \phi^{g}=\eta_f \pi
\begin{pmatrix}
1\\
1
\end{pmatrix}.
\end{align}
Here, $\eta_f=0$ and $1$ correspond to trivial, and the sign change transformations.
This $\mathbb{Z}_2^{f}$ always exists in fermionic SPT phase.

One can consider symmetry of fermion bilinear $G$; the total symmetry is 
$G_f=G\times \mathbb{Z}_2^{f}$.
Since $G$ is a bosonic symmetry, for a given bosonic SPT phase, 
one can always construct a fermionic SPT phase.
The symmetry transformation $g$ and its transformation in 
fermionic SPT phases are determined in a way similar  
to bosonic SPT phases, (\ref{cond 1}) and  $g^N=e$,
except that now the identity in $g^N=e$
allows both the trivial and the sign change transformation. 
For $G_f=\mathbb{Z}_2\times \mathbb{Z}_2^{f}$, 
let $g$ be the generator of the $\mathbb{Z}_2$ symmetry transformations.
It acts on the edge theory as
 \begin{align}
&&(W^g)^2=\delta_{IJ}, ~~(W^{g})^{T}KW^{g}=K,\cr
&&(I+W^{g})\delta \phi^{g}=\eta\pi
\begin{pmatrix}
1 \\
1
\end{pmatrix}
~~mod~~2\pi.
\end{align}
This system is a SPT phase when 
\bea
\eta=0,
\eea
and $t_1-t_2$ mod 2 with ${\bf t}=(t_1,t_2)$ a charge vector.
Other cases are trivial.
Especially any phase with $t_1=t_2$ $(q=1)$ is trivial.
As we will show below, 
the charged topological entanglement entropy is calculated as 
\bea
\mathscr{C}= ({1-q^2}),
\label{fermion SPT result}
\eea
which indeed is zero/non-zero for trivial/topological phases.

\subsubsection{Charged topological entanglement entropy}

As in the bosonic case, the entanglement chemical potential generates a Wilson loop along the entangling surface
\bea
W(\mue)=\exp\left(-\mue \oint_{r=0} (a^1 - q a^2)\right),
\eea
whereas the entanglement flux generates a Wilson loop around the entangling surface
\bea
W(\phie)=\exp\left(\phie \oint_{r=1} (a^1 - q a^2)\right).
\eea
From the propagator (\ref{a-prop}), one can expect that the topological
term in the presence of the entanglement chemical potential and the flux gives
\bea
\langle W(\mue)W(\phie)\rangle= \exp \left(   2\pi i (1-q^2) {\mue} {\phie } \right),
\label{CTEEfermi1}
\eea
from which we conclude the charged topological entanglement entropy \eqref{fermion SPT result}.

One can derive the same result from the CFT point of view.
The entanglement hamiltonian $H_E$ and the entanglement charge $Q_E$ are given by
\bea
H_E={1\over 2\pi}\int dx {1\over 2}((\pd_x \phi_1)^2+(\pd_x \phi_2)^2),
\eea
and
\bea
Q_E={1\over 2\pi}\int dx(k_2 (\pd_x \phi_1)-l_2(\pd_x\phi_2)).
\eea
The edge modes $\phi_1$ and $\phi_2$ are already in the chiral and the anti-chiral modes.
The global $\mathbb{Z}_N $ symmetry is encoded into the boundary conditions of the fields
\bea
\phi_1(x+2\pi)&=&\phi_1(x)+2\pi(n+{k_1\over N}),\cr
\phi_2(x+2\pi)&=&\phi_2(x)+2\pi (m+{l_1\over N}),
\eea
which determines the mode expansions
\begin{align}
\phi_1(x,t)&=\phi_1^0+p_1(t+x)+i\sum {a_n\over n}e^{-in(t+x)},\cr
\phi_2(x,t)&=\phi_2^0+p_2(t-x)+i\sum {b_n\over n}e^{-in(t-x)}.
\end{align}
The momenta are
\begin{align}
p_1&=n+{k_1\over N},
\quad
p_2=-(m+{l_1\over N}). 
\end{align}
We interpret 
{$k_1/N$}  and 
{$l_1/N$} as the entanglement flux $\phie$.
The CFT partition function is
\begin{align}
& 
Z(\tau)
=
\eta(\tau)\sum_{n,m=-\infty}^{\infty}
\nonumber \\
&\quad 
\times 
\exp\Big[-\pi \tau_2(n+{k_1\over N})^2-\pi \tau_2 (m+{l_1\over N})^2
\nonumber \\
&\qquad 
+2\pi i\mue k_2(n+{k_1\over N})
-2\pi i \mue l_2 (m+{l_1\over N})\Big]. 
\end{align}
By performing the
Poisson resummation with respect to $n$, one obtains
\begin{align}
Z(\tau) &=
\sqrt{{1\over \tau_2}}\eta(\tau)\sum_{n,m=-\infty}^{\infty}
\nonumber \\
&\quad 
\times
\exp\Big[
-{\pi\over \tau_2}(\mue k_2-n)^2+2\pi i {k_1\over N}n
\nonumber \\
&\qquad 
-\pi\tau_2(m+{l_1\over N})^2-2\pi i\mue  l_2(m+{l_1\over N})
\Big].
\end{align}
The S transformation $\tau_2\to 1/\tau_2$ and the change of parameters
\bea
\begin{split}
&m\to n, n\to -m,
l_1\to -\mue N k_2,
l_2\to -{k_1\over \mue N}, \\
&k_1\to N\mue l_2,
k_2\to {l_1\over N\mue}.
\end{split}
\eea
give
\begin{align}
	&
Z(\tau)_{k_1,k_2,l_1,l_2}
\nonumber \\
&
=Z(-1/\tau)_{l_2,l_1,-k_2,-k_1}\exp\Big[
2\pi i {\mue\over N}(k_1 k_2-l_1 l_2)
\Big].
\end{align}

In the thermodynamical limit $\tau_2\to 0$ limit, $Z(1/\tau_2)$ gives only the diverging term ($\eta(\tau)\sqrt{{r\over \tau_2}}$) and the constant term gives
\bea
\log Z=
\exp\left[
2\pi i {\mue\over N}(k_1 k_2-l_1 l_2)
\right].
\eea 
Therefore, setting $k_1/N=\phie, l_1/N=q\phie$ and $k_2=1,l_2=q$ 
the charged topological entanglement entropy is
\bea
\gamma_{c}= ({1-q^2}) (2\pi i \mue \phie).
\label{CTEEfermi2}
\eea
This result agrees with (\ref{CTEEfermi1}).

As in the bosonic case,
once we know the charged entanglement entropy, one can distinguish different SPT phases
based on the analysis in \cite{Cheng Gu}.
In the case of fermionic phases, the braiding of a fermion generates a phase which is a multiple of $\pi/N$.
The total phase factor of excitations is a summation of the contributions from the vortices and the fermions.
Whether the phase factor from a fermion gives an additional constraint or not depends on whether $N$ is odd or even.
As demonstrated in \cite{Cheng Gu},
for odd $N$, $\mathscr{C} $ is identified modulo $N$, therefore the classification is $\mathbb{Z}_N$
while for  even $N$, $\mathscr{C} $ is identified modulo $2N$ which results in $\mathbb{Z}_{2N}$
classification.

\section{Shifted Entanglement Entropy}
\label{Shifted bulk story}

Since the momentum polarization also twists the wave function, one can consider a shifted topological entanglement entropy.
From the generalized entanglement entropy point of view,
the momentum polarization generates a dislocation along the entangling surface.
When one moves around the entangling surface, one's position is shifted by the amount of the Burgers vector.
This is described by the co-frame field.

The action for a fermion model coupled to a frame field is
\bea
S=\int d^3x\, \det(e)\bar{\psi}(\gamma^{a}e^{\mu}_a\pd_{\mu}-m)\psi.
\label{torsion free fermion}
\eea
By integrating the massive fermion, one obtains \cite{torsion}
\bea
S={1\over 32\pi}I_{T}(m)\int
{d^3x\,}
\epsilon^{\mu\nu\rho}e^{a}_{\mu}\pd_{\nu}e^{b}_{\rho}\eta_{ab},
\eea
where
\bea
I_T(m)=\int_{0}^{\infty} dy y{\pd\over \pd y}{m\over (y+m^2)^{1/2}}.
\eea
As discussed in \cite{torsion},  this coefficient can be interpreted as the
Hall viscosity with an appropriate regularization.

One can bosonize the fermion current in the same way as in the previous section.
The partition function of the fermions coupled with an external coframe field $E^{ex}$ is
\bea
Z[E^{ex}]=\int \mathcal{D}[\bar{\psi},\psi] \exp\left(- S[\bar{\psi},\psi,E^{ex}]  \right).
\eea
This action is invariant under a shift transformation $Z[E^{ex}]=Z[ E^{ex}+e]$ for a torsion free frame
$T^{I}_{\mu\nu}=\pd_{(\mu}e^{I}_{\nu)}=0$.
Then by introducing a Lagrangian multiplier $(d-2)$ form field $b^{I}_{\mu\nu\cdots}$ with a Lorentz index $I$
the action can be rewritten
\bea
\begin{split}
Z[E^{ex}]=&\int \mathcal{D}[e,b] Z[E^{ex}+e]\\
&\times \exp\left( -{i} \int b^{I}\wedge T^{J}(e)  \eta_{IJ}
\right),
\end{split}
\eea
where $\eta_{IJ}=\text{diag}(1,1,1)$.
By shifting $e\to e-E^{ex}$, we obtain
\begin{align}
&Z[E^{ex}]=\int \mathcal{D}[e,b] Z[e]
\nonumber \\
&\quad 
\times \exp\left( -{i} \int b^{I}\wedge\left(T^{J}(e)
-T^{J}(E^{ex})
\right)
\eta_{IJ}
\right).
\end{align}
The bosonization rule for the {momentum current}
is
\bea
T^{I}={1\over i}{\delta \log Z[E^{ex}]\over \delta E^{I,ex}}=*db^{I}.
\eea
In the case of the free fermion
(\ref{torsion free fermion}), the current is
\bea
T^{I}_{\mu}=\bar{\psi}p^{I}\gamma_{\mu}\psi.
\eea
The effective action is
\begin{align}
S=-i \int b^{I}(de^{J}-dE^{J,ex})\eta_{IJ}+
{i \zeta_{H}\over 2}\int e^{I}d e^{J}\eta_{IJ}.
\end{align}
In the path integral picture, the entanglement shift flux ($\alpha^{I}_E$) and the entanglement shift potential ($\beta^{I}_E$) are described by the background co-frame fields
\bea
E^{I,ex}_{y}=\alpha^{I}_E,~~~~E^{I,ex}_{\theta}=\beta^{I}_E.
\eea
These two terms generate co-frame field Wilson loops
\begin{align}
&
W(\alpha^{I}_E)=\exp\left( i\alpha^{I}_E \oint_{r=1}b^{I}_{\theta}d\theta\right) ,
\nonumber \\
\quad
\mbox{and}
\quad 
&
W(\beta^{I}_E)=\exp\left( -i \beta^{I}_E \oint_{r=0}b^{I}_{y}dy\right).
\label{b-wilson}
\end{align}
By 
integrating $e^{I}$ out, one obtains the Chern-Simons action for $b^{I}$.
Then the partition function in the presence of the Wilson loops (\ref{b-wilson}) for $b^{I}$ is
\bea
\left\langle W(\alpha^{I}_E) W(\beta^{J}_E)\right \rangle= \exp
\left({i\zeta_{H}} \alpha^{I}_E \beta^{J}_E\eta_{IJ} \right).
\eea
One can conclude that the universal part of the shifted entanglement entropy, {\it the shifted topological entanglement entropy} $\gamma_{s}$,  is
\bea
\gamma_{s}={ \zeta_{H}} (i\alpha^{I}_E \beta^{J}_E\eta_{IJ}).
\eea
The shifted topological entanglement entropy is non-zero and the coefficient of $(i \alpha^{I}_E \beta^{J}_E\eta_{IJ})$ contains 
physical information: the Hall viscosity.
There are a couple of differences between the charged entanglement entropy and the shifted entanglement entropy.
In the case of the charged entanglement entropy, the coefficients of the Chern-Simons theory, i.e., components of the $K$ matrix, are quantized in integers.
On the other hand, the coefficient of the coframe field Chern-Simons theory, which is the Hall viscosity $\zeta_{H}$, is not quantized given that the gauge group corresponding to diffeomorphism is non-compact. Therefore, the shifted topological entanglement entropy is not quantized.
Another point is the symmetry.
In the charged case, one could introduce multiple external and internal $U(1)$ gauge fields.
For instance, for SPT phases with $\prod_{i=1}^{M} \mathbb{Z}_{N_i}$ symmetry
described by $K_{IJ}$ matrix,
one could introduce external fields $A^{ex}_i$ ($i=1,\cdots, M$) as well as internal fields $a^{I}$.
In the shifted case, 
the co-frame fields couple with all the degrees of freedom equally, 
and therefore
it is natural to have only one type of external and internal co-frame fields $E^{I,ex}$ and $e^{I}(b^{I})$
especially in interacting systems.


\section{Concluding Remarks}

In this paper, we apply the grand canonical entanglement entropies to study two dimensional symmetry protected topological phases with onsite unitary symmetries.
 The topological entanglement entropy for SPT phases always vanishes and it does not distinguish topological phases from trivial ones. 
Our main observation is that the universal part of these grand canonical entanglement entropies are sensitive to distinguish non-trivial short
range entangled topological phases. Therefore, they play a similar role as the topological entanglement entropy in topologically ordered (long range entangled)
 states.
 The inclusion of the entanglement potentials effectively plays a role of insertion of Wilson loops along and around the entangling surface.
 Therefore, one can interpret the charged topological entanglement entropy as measuring a braiding statistics of quasi-particles.
 In higher dimensional topological phases, quasi-particles can be replaced by loops and other higher dimensional objects.
 One difference between the ordinary topological entanglement entropy and the charged topological entanglement entropy is 
 that the topological entanglement entropy is a  real number while the charged topological entanglement entropy is imaginary
 reflecting the fact that it is a phase factor of the partition function.
The topological information is encoded in the coefficient $\mathscr{C}$ of $(2\pi i \mue\phie)$ in $\gamma_c$.
Its identification is determined by a physical reasoning: 
 In the case of the Chern Insulator, $\mathscr{C}$ is identical to the Chern number, while for
SPT phases, it is the Berry phase of the braiding operation.

 We did not address the time reversal symmetry in this paper. 
 The main obstacle is that at this point there is no known gauge field that generates the time reversal symmetry. And the chemical potential 
 associated with it is not clear.
 We hope to come back to this issue in the future.

\acknowledgments
We are grateful to A.~Furusaki, A.~Kapustin, T.~Morimoto, T.~Neupert, T.~Okuda, M.~Oshikawa, M.~Rudner, M.~Sato, and T.~Takayanagi for useful discussions. 
SM also wishes to acknowledge Caltech, ISSP, the University of Tokyo, and Riken
for their hospitality.
This work was supported in part by the National Science Foundation grant DMR-1455296 (XW and SR) at the University of Illinois, 
and by Alfred P. Sloan foundation.



\begin{thebibliography}{99}
\bibitem{caution on RBH}
There is a caveat about the rotating black holes.
Since the horizon is non-compact, the black hole with a real angular potential has a singularity.
This can be understood intuitively from the fact that one cannot consider
a thermodynamics of a uniformly rotating space.
As we go away from the center of the rotation, at some point, the speed of the rotation exceeds the speed of light.
Therefore the partition function of this system at finite temperature diverges.
On the other hand, one could still consider an imaginary angular potential.
\end{thebibliography}
\end{document}